\documentclass[12pt]{article}
\usepackage{amsmath}
\usepackage{epsfig}
\begin{document}
\title{Role of atomic electron shell in the double beta decay}
\author{E. G. Drukarev$^{1}$, M. Ya. Amusia$^{2,3}$, L. V. Chernysheva$^{3}$\\
{\em $^{1}$National Research Center "Kurchatov Institute"}\\
{\em B. P. Konstantinov Petersburg Nuclear Physics Institute}\\
{\em Gatchina, St. Petersburg 188300, Russia}\\
{\em $^2$ Racah Institute of Physics, The Hebrew University}\\
{\em Jerusalem 91904 Jerusalem, Israel}\\
{\em $^3$ A. F.Ioffe Physical-Technical Institute,}\\
{\em St. Petersburg 194021,Russia }}
\date{}
\maketitle

\begin{abstract}
{We demonstrate that the limiting energy available for ejected electrons in double beta decay is diminished by about
400 eV due to inelastic processes in the atomic electronic shell.}
\end{abstract}

\section{Introduction}

The double beta decay $(2\nu\beta\beta)$ has been observed for eleven nuclei \cite{1}-\cite{3}. There are 35 nuclei for which the $\beta$ decay is forbidden while the double beta decay
\begin{equation}
(A,Z)\rightarrow (A, Z+2)+2e^-+2\bar\nu_e
\label{1}
\end{equation}
can take place \cite{4}. Several attempts to detect the neutrinoless double beta decay $(0\nu\beta\beta)$
\begin{equation}
(A,Z)\rightarrow (A, Z+2)+2e^-
\label{2}
\end{equation}
as well as the future projects in this direction are described in the review \cite{4}. Observation of the neutrinoless decay would mean that the electron neutrino is a Majorana particle which coincides with its antiparticle, i.e. $\nu_e \equiv \bar\nu_e$. In the process $(2\nu\beta\beta)$ neutrinos carry the energy $M_{N}(Z+2)-M_N(Z)-E$ with $E$ the energy of ejected  electrons while
$M_N(Z+2)$ and $M_N(Z)$ are the masses of the nuclei $(A,Z+2)$ and $(A,Z)$. 
A sign of the neutrinoless process $(0\nu\beta\beta)$ would be the ejection of the electrons with the energy $E=M_{N}(Z+2)-M_N(Z)$ leaving no room for the neutrinos. As it stands now, the neutrinoless decay have not been detected \cite{4}.

The actual experiments are carried out for the decays of atoms but not of the bare nuclei. It was claimed in the preprints \cite {5}-\cite{7} that this inserts an uncertainty of the order of $3$ keV to the analysis. It was claimed also \cite{7} that the neutrinoless decay $(0\nu\beta\beta)$ actually have been detected in experiments \cite{8} and \cite{9}. This stimulated us to analyze the problem of the role of  the atomic electronic shell in the double beta decay. We employ the relativistic system of units with $\bar h=c=1$.

\section{Double beta decay of the atom}

Now we turn to the decay $(2\nu\beta\beta)$ of the atom ${\cal A}_Z$ with the nucleus of the charge $Z$.
We assume the atom ${\cal A}_Z$ to be in its ground state. In the main mode of the decay
\begin{equation}
{\cal A}_Z\rightarrow {\cal A}_{Z+2}^{++}+2e^-+2\bar\nu_e,
\label{3}
\end{equation}
the double charged positive ion ${\cal A}_{Z+2}^{++}$ is also in the ground state. The mass of the atom
${\cal A}_Z$ and that of the ion ${\cal A}_{Z+2}^{++}$ in the ground state are $M_{At}(Z)=M_N(Z)+E_b(Z)$,and
$M_{At}^{(0)}(Z+2)=M_N(Z)+E_b^{(0)}(Z+2)$ with $ E_b(Z)$ and $E_b^{(0)}(Z+2)$ the total energies of the atomic electronic shells.
The upper index $(0)$ labels that the ion ${\cal A}_{Z+2}^{++}$ is in the ground state. Thus the largest energy available for the ejected electrons is
\begin{equation}
Q^{(0)}=Q_N+E_b(Z)-E_b^{(0)}(Z+2),
\label{4}
\end{equation}
with $Q_N=M_N(Z)-M_N(Z+2)$ the largest electron energy available in the decay of the nucleus presented by Eq.(\ref{1}). One can see that $Q^{(0)}>Q_N$, i.e. the electronic shell becomes more bound. It transfers the energy to the ejected electrons. The masses of the atom $M_{At}(Z)$ and of the ion $M_{At}^{(0)}(Z+2)$ can be measured with good accuracy. Thus the limiting energy $Q^{(0)}$ is a well established observable.

The electronic shell can be in an excited state after the decay. This can be an excited state of the ion ${\cal A}_{Z+2}^{++}$. Also some of $Z$ bound electrons can be moved to continuum. This shifts the observable value of $Q^{(0)}$
to
\begin{equation}
Q=Q^{(0)}-\delta Q.
\label{5}
\end{equation}
To obtain $\delta Q$ introduce the excitation energy $E^*_{n0}=E^{(n)}(Z+2)-E^{(0)}(Z+2)>0$.
Denoting the differential distribution of the double beta decay with excitation of the atomic electronic shell to the state $n$ as $dW_n/d\Gamma$ we can present
\begin{equation}
\delta Q=\sum _n E_{n0}^*\frac{dW_n/d\Gamma}{dW_0/d\Gamma},
\label{6}
\end{equation}
with $\sum_n$ meaning the sum over the states of the discrete spectrum and integration over the continuum states.
The energy of the state $n$ should not exceed the limiting energy $Q^{(0)}$.

Note that in the process $(2\nu\beta\beta)$ the  energy $Q^{(0)}$ makes several MeV \cite{4}. A simple analysis based on estimation of the phase volume shows that both ejected electrons should be fast. Thus their velocities (in units of $c$) are of the order of unity. The atomic velocities are of the order $\alpha Z^{1/3}$, i.e. they are much smaller. This enables us to employ the shake off (SO) approximation in which the final state interactions between the beta electrons and atomic electronic shell are neglected \cite{10}. In this approach the amplitude for the decay in which the atomic shell transfers to the state $n$ is
\begin{equation}
F^{(n)}=F_N\langle \Phi_n|\Psi\rangle.
\label{7}
\end{equation}
Here $F_N$ is the amplitude for the nuclear decay, $\Psi$ and $\Phi_n$ are the wave functions of $Z$ electrons
in the ground state of the field of the nucleus with the charge $Z$ (the atom) and in the state $n$ of the field of the nucleus with the charge $Z+2$. This provides
\begin{equation}
\delta Q=\sum _n E_{n0}^*|\langle \Phi_n|\Psi\rangle|^2.
\label{8}
\end{equation}
Introducing the total change of the electronic shell energy
$E_{n0}=E^{(n)}(Z+2)-E_b(Z)>0$ and presenting
$E^*_{n0}=E_{n0}+E_b(Z)-E^{(0)}(Z+2)$, we can write $\delta=\delta_1+\delta_2$
with
\begin{equation}
\delta Q=\delta_1+\delta_2;  \quad
\delta_1=(E_b(Z)-E^{(0)}(Z+2))\sum_n|\langle \Phi_n|\Psi\rangle|^2 ;
\label{9}
\end{equation}
$$\delta_2=\sum _n E_{n0}|\langle \Phi_n|\Psi\rangle|^2.$$

Since Eq.(\ref{9}) contains only the differences of the energies we can write it in terms of the binding energies
subtracting the mass terms, i.e. putting $E_b(Z)=m_eZ+\varepsilon_b(Z)$,   $E^{(0)}=m_eZ+\varepsilon^{(0)}(Z)$ ,
$E^{(n)}(Z+2)=m_eZ+\varepsilon^{(n)}(Z+2)$ with $m_e$ the electron mass at rest and $E_{n0}=\varepsilon_{n0}$.
Thus the two last equalities of Eq.(\ref{9}) can be presented as
\begin{equation}
\delta_1=(\varepsilon_b(Z)-\varepsilon^{(0)}(Z+2))\sum_n|\langle \Phi_n|\Psi\rangle|^2 ; \quad
\delta_2=\sum _n \varepsilon_{n0}|\langle \Phi_n|\Psi\rangle|^2.
\label{9a}
\end{equation}

The squared SO matrix element $|\langle \Phi_n|\Psi\rangle|^2$ drops as $\varepsilon_n^{-4}$ if the excitation energy $
\varepsilon^*_{n0}$ exceeds strongly the ground state energy $|\varepsilon^{(0)}|$ \cite{11}. Thus the sums over $n$
on the right hand side of Eqs.(\ref{6})and (\ref{9}) are saturated at
$\varepsilon^*_{n0} \sim |\varepsilon_0| \ll
Q^{(0)}$.
Hence we can assume that the sum over $n$ is carried out over all states with $\varepsilon_{n0}>0$.
If the atom is treated as a nonrelativistic system, they compose  a closed set of states. This enables us to employ closure.
This provides
\begin{equation}
\delta_1=\varepsilon(Z)-\varepsilon^{(0)}(Z+2),
\label{10}
\end{equation}
while $\delta_2=\langle \Psi|H(Z+2)-H(Z)\Psi\rangle$ with $H(Z)$ and $H(Z+2)$ the Hamiltonians of $Z$ electrons in the fields of the nuclei with the charges $Z$ and $Z+2$. Thus we obtain \cite{11}
\begin{equation}
\delta_2=\langle \Psi|\sum_{k \leq Z}\Big(\frac{-2\alpha}{r_k}\Big)|\Psi\rangle,
\label{11}
\end{equation}
with the sum carried out over all electrons bound in the atom ${\cal A}_Z$. Both $\delta_1$ and $\delta_2$  can be calculated with high accuracy. Note that Eq.(\ref{9}) presents the shift $\delta E$ as a difference of two large values.

Corrections to the SO approximations can be obtained by inclusion of the final state interaction (FSI) between the beta electron and the bound ones. We can write
\begin{equation}
\delta Q=\delta_1+\delta_2+\delta_{FSI}.
\label{12}
\end{equation}
The leading correction is proportional to the squared Sommerfeld parameter $\xi$ of the beta electron moving with the velocity $v$ which is $\xi^2=\alpha^2/v^2$ with $\alpha=1/137$ the fine structure constant. The expressions describing probability for transition of the atomic shell to any excited state with inclusion of the leading FSI terms are presented in \cite{11}. Since the latter drop as $\varepsilon_n^{-2}$ at large $\varepsilon_n \gg |\varepsilon_0|$, one can not calculate the FSI contribution $\delta_{FSI}$ to the shift $\delta E$ by using the closure condition. Assuming that the energy transferred to the bound electrons does not exceed certain value $\varepsilon_{max}$ we can estimate \cite{11}
\begin{equation}
\delta_{FSI}=\xi^2\sum_k
\frac{a_k\langle \Psi|r_k^{-2}|\Psi\rangle}{m}\ln{\frac{\varepsilon_{max}}{|\varepsilon_k|}},
\label{13}
\end{equation}
with $a_k$ the number of electrons in the bound state $k$. Note that if the atomic electrons obtain the energy $\varepsilon_{max}$, the energy of the beta electrons can not exceed the value $E=Q^{(0)}-\varepsilon_{max}$.

\section{Numerical results}

Now we carry out numerical calculations. In actual computations the atom is presented as a system of electrons described
(at least in the first step) by single-particle functions. Note that we can employ only the nonrelativistic functions since the positive-energy states compose the closed system only in nonrelativistic case. We use our Hartree-Fock computer codes \cite{12}

Start with the double beta decay of germanium($Z=32$)\cite{8}, \cite{4}.
We find $\varepsilon(Z)=-56449.2$eV and
$\varepsilon^{(0)}=
- 65246.3$eV for the ground state energies of the atom of Ge and of the ion Se$^{++}$ ($Z=34$).
This provides $\delta_1=8797.1$eV. We obtain also $\delta_2=-8446.1$ eV. This provides $\delta_1+\delta_2=351$eV.
Note that the value of $\delta_1$ calculated in relativistic approach (Hartri-Fock-Dirac approximation) is about $2\%$ larger. If the relative size of relativistic correction to $\delta_2$ is of the same order as that to $\delta_1$,
we find that relativistic corrections to $\delta Q$ are also of the order $2$\%.
The small magnitude of relativistic effects makes the nonrelativistic calculation reasonable.

In calculation of the FSI contribution $\delta_{FSI}$ we can put $\xi^2=\alpha^2$ since $Q^{(0)} \approx 2$ MeV. Assuming $\varepsilon_{max} =3$ keV we find that $\delta_{FSI}=0.6$eV. Thus we obtained
\begin{equation}
\delta Q =352 eV.
\label{14}
\end{equation}

For the double beta decay of xenon($Z=54$) \cite{2}, \cite{3}, \cite{13}
$\varepsilon(Z)=-196714.2$eV while for the ion Ba$^{++}$ (Z=$56$) $\varepsilon^{(0)}=
- 214419.3$eV. Thus $\delta_1=17705.1$eV. On the other hand  $\delta_2=-17292.4$ eV. This provides
$\delta_1+\delta_2=413$eV. The FSI contribution is $\delta_{FSI}=1$eV. Thus we obtain
\begin{equation}
\delta Q =414 eV.
\label{15}
\end{equation}
The value of $\delta_1$ calculated in relativistic Hartree-Fock-Dirac approach provides the value which is about $1$ keV larger. If the relative difference of the Hartree-Fock -Dirac and Hartree -Fock results for the value $\delta_2$ is of the same order of magnitude as for $\delta_1$ the uncertainty of our nonrelativistic result is of the order $10$  percent.

\section{Summary}

We carried out nonrelativistic calculation for the shift of the limiting energy available for the ejected electrons in double beta decay caused by inelastic processes in electronic shell. We demonstrated that the energy diminishes by about $400$ eV. We estimated the accuracy of our calculations. Our result does not alter the earlier conclusions  that the neutrinoless mode have not been observed yet.

We thank Ya. I. Azimov and Yu. N. Novikov for stimulating discussions.

\end{document}